\documentclass[12pt,A4]{article}
\usepackage[cp866]{inputenc}
\usepackage[russian]{babel}
\usepackage{epsfig}
\usepackage{array}
\usepackage{amsmath}
\usepackage{amssymb}
\usepackage{indentfirst}

\newcommand{\m}{\mathbf}
\newcommand{\be}{\begin{equation}}
\newcommand{\ee}{\end{equation}}
\newcommand{\bea}{\begin{eqnarray}}
\newcommand{\eea}{\end{eqnarray}}

\begin{document}
\title{
TOWARDS EXPERIMENTAL SEARCH FOR THE ATOMIC EFFECT IN $\alpha$ DECAY: COMPARISON OF THE ADIABATIC APPROACH WITH THE FROZEN SHELL MODEL}

\author{F. F. Karpeshin$^{1)}$, M. B. Trzhaskovskaya $^{2)}$}

\maketitle

$^{1)}$ Mendeleev All-Russian Research
Institute of Metrology,  Saint-Petersburg, Russia

$^{2)}$ Petersburg Nuclear Physics Institute, Kurchatov Research Center, \\ Gatchina, Russia

\renewcommand\abstractname{}
\abstract{We study the important for the experiment features of the effect of the electron shell on $\alpha$ decay in the adiabatic approach. The effect is  of tenths of a percent or less in magnitude, depending on the transition energy and the atomic number. On the other hand, we show the predominant role of the inner shells:  more than 80 \% of the effect are due to the  $1s$ electrons. This fact is crucial for the experiment, allowing one to perform measurements  in the same storage rings, comparing for example, the probability of decay in the bare nuclei and Helium-like ions. The analysis is presented concerning the reasons for the relative success and the limits of applicability of the model  of  ``frozen'' electron shell used to calculate the effect for more than half a century. }

\clearpage
\begin{center}
{\large\bf
ЭКСПЕРИМЕНТАЛЬНЫЕ АСПЕКТЫ АДИАБАТИЧЕСКОГО ПОДХОДА В ОЦЕНКЕ ВЛИЯНИЯ ЭЛЕКТРОННОГО ЭКРАНИРОВАНИЯ НА АЛЬФА-РАСПАД} \\

\bigskip
\bigskip
Ф.Ф.Карпешин$^{1)}$, М.Б.Тржасковская$^{2)}$    \\
\bigskip
$^{1)}$ ВНИИ Метрологии имени Д.И.Менделеева

$^{2)}$ НИЦ ``Курчатовский Институт'', ПИЯФ имени Б.П.Константинова
\end{center}

\abstract{
Исследуются важные для эксперимента особенности влияния электронной оболочки на $\alpha$-распад в рамках адиабатического подхода. Величина эффекта составляет десятые доли процента и меньше, в зависимости от энергии перехода и атомного номера. С другой стороны, показана доминирующая роль внутренних оболочек: более 80\% эффекта обязаны $1s$-электронам. Это обстоятельство играет решающую роль для эксперимента, позволяя производить измерения малого эффекта разностным методом в одних и тех же накопительных кольцах, сравнивая  например, вероятности распада в голых ядрах и гелие-подобных ионах. Проанализированы причины относительного успеха и пределы применимости  модели  ``замороженной'' электронной оболочки, применявшейся для расчета эффекта на протяжении более полувека.}

\normalsize
\section{Введение}

$\alpha$ распад и его торможение электронной оболочкой суть явления, происходящие на разных структурных уровнях микромира, характеризующихся различной шкалой как по линейным размерам, так и по массе участвующих частиц, их энергии, по силе их взаимодействий. Они управляются силами разной природы: сильным  ядерным взаимодействием, область действия которого ограничена ядерными размерами, и гораздо более слабым кулоновским взаимодействием электронов друг с другом и $\alpha$-частицей, но влияние которого сказывается на гораздо больших расстояниях. {\it Apriori,} сильного влияния электронной оболочки на альфа-распад, при энергии альфа-частицы несколько МэВ, трудно ожидать. Влияние других факторов ядерного происхождения, например деформации ядра,  несравнимо сильнее. Тем не менее, разницу в периодах альфа-распада голого ядра и нейральных атомов можно измерить в эксперименте, что представляет большой интерес с точки зрения тестирования теории Гамова \cite{kar}.
Следует принять во внимание, что перестройка оболочки атома приводит к уменьшению конечной энергии альфа-частицы на величину $\Delta Q \approx$  40 кэВ в тяжелых атомах --- типичных альфа-распадчиках. Это уже приводит к торможению распада в буквальном смысле слова, в противоположность тому, что можно было бы ожидать исходя из притяжения между альфа-частицей и электронами, хотя и не сказывается на динамике самого альфа-распада. В то же время последовательный учет этого обстоятельства, ввиду перечисленных выше факторов, стал камнем преткновения во многих ранних расчетах. Первые попытки прямой подстановки величины $\Delta Q$ в формулу для проницаемости кулоновского барьера привели к завышенным оценкам \cite{PR1}. Удивительно, что на протяжении десятилетий расчеты выполнялись в модели ``замороженной'' элетронной оболочки (ЗО), суть которой прямо противоположна тому, что сказано выше (\cite{perl,PR2,PRZin,pol} и ссылки там). По-видимому, только в работах \cite{perl,PRZin} был сделан решающий шаг  к  адекватной физической трактовке явления, хотя и в рамках той же модели. Логическая точка в развитии модели ЗО была поставлена в работе \cite{pol}, в которой были выполнены наиболее последовательные расчеты альфа-распада на примере цепочки атомов радона. Дальнейшие попытки эксплуатации этой модели в погоне за гипотетическими эффектами влияния оболочки на конечную вероятность альфа-распада после прохождения барьера \cite{chu}  уже противоречат здравому смыслу, поскольку за барьером, принимая во внимание регулярный (квазиклассический, ср. \cite{dzub})  характер $\alpha$-атомного потенциала, ток распадых альфа-частиц в любом случае сохраняется. Впрочем, математически постановка задачи в работе \cite{chu}  не сформулирована. Мы еще вернемся к этому  пункту при обсуждении физических аспектов проблемы.

    Указанные выше физические особенности задачи адекватно отражаются  термином ``адиабатичность''.
Задача о прохождении альфа-частицы через электронную оболочку вполне аналогична задаче о распределении мюонов между осколками мгновенного деления в мюоном атоме урана \cite{ober, lea,risse,dem}.
В последнем случае, деление ядра происходит в присутствии мюона на $K$-орбите. Мюон много легче осколков. В свою очередь, скорость последних много меньше скорости мюона на орбите. Это приводит к формированию адиабатических квазимолекулярных состояний в процессе формирования и разлета осколков. Именно перестройка квазимолекулярного состояния приводит к наблюденному в эксперименте эффекту подавления вероятности мгновенного деления вследствие повышения барьера деления \cite{lea,risse}. В результате выход деления уменьшается на порядок по сравнению с обычным делением, индуцированным нейтронами или фотонами. Неучет данного эффекта привел к задержке открытия мгновенного деления на добрый десяток лет.  В качестве второго примера можно указать нильссоновские орбитали и их влияние на равновесную деформацию ядра \cite{berl}.

    Альфа-распад можно рассматривать как предельный случай деления с большой асимметрией осколков. Подобно мюону, электронная оболочка образует квазимолекулярные состояния в поле двух центров. Электронные термы играют роль потенциальной энергии взаимодействия осколков. Именно такая картина для описания эффекта влияния электронной оболочки на альфа-распад была предложена в работе \cite{kar}. Расчеты, выполненные для ряда ядер, показали, что последовательный учет адиабатических эффектов приводит к уменьшению вероятности распада по крайней мере в несколько раз по сравнению с моделью ЗО. Эти идеи подтверждаются,  развиваются и применяются к расчету сечений реакции синтеза (обратной к альфа-распаду) ядер, имплантированных в металлическую матрицу, в работе \cite{dzub}. Реализуется предсказанное в  \cite{kar} расхождение с прежней теорией, основанной на модели  ЗО.
В настоящей работе в последующих разделах мы остановимся на физических аспектах явления, важных для постановки эксперимента. Экспериментально вероятность распада может быть легко померена например, в накопительных кольцах на ускорителях, подобных имеющимся в GSI или Ланчжоу. В Заключении подведем итоги, сформулируем перспективы экспериментальных исследований.

\section{Физические предпосылки }

      Обычно в теории $\alpha$-распада выделяют три фактора: вероятность формирования кластера внутри ядра, частота его столкновений со стенкой и проницаемость кулоновского барьера. Влиянием электронной оболочки на первый фактор пренебрегаем.  Второй фактор рассматривается как подгоночный параметр (ср. \cite{den,sochi}). Рассмотрим проницаемость барьера:
\be
P = e^{-2S}\,.      \label{Pal}
\ee
В (\ref{Pal}) действие (например, \cite{fur}) для голого ядра
\be
S = \int_{r_1}^{r_2}\sqrt{2m(U_C(R) - Q_\alpha)}\ dR\,. \label{BN}
\ee
В выражении (\ref{BN}) $r_1$, $r_2$ --- классические точки поворота, $m$ --- приведенная масса  $\alpha$-частицы, $U_C(R)$ --- кулоновская энергия отталкивания ядер.

      В  случае нейтрального атома действие (\ref{BN}) переходит в
\be
S_{at} = \int_{r_1'}^{r_2'}\sqrt{2m(E(R ) -Q_\alpha)}   \,, \label{eq17}
\ee
где  $E(R)$ --- терм двуцентровой квазимолекулы \cite{lan}. Он включает в себя, помимо энергии отталкивания ядер, и электронную часть $\epsilon(R)$ --- мгновенное значение электронного гамильтониана $h(R)$:
\be
E(R) = U_C(R)+ \epsilon (R)\,.  \label{eq18}
\ee
Модифицируя потенциальную энергию,  электронная энергия изменяет и точки поворота, новые значения которых становятся равны $ r_1'$   и  $r_2'$.

      На малых расстояниях электронная часть энергии терма отрицательна и, очевидно, максимальна в материнском ядре, то есть при $R\to 0$. Поэтому, когда альфа-частица проходит путь от ядра к барьеру и дальше сквозь него на бесконечность, ей приходится как бы ``идти в горку'', преодолевая подъем.   На это расходуется часть ее энергии $\Delta Q$, приводя к уменьшению $Q_\alpha$ в нейтральном атоме. Отсюда следует, что учет электронной оболочки замедляет альфа-распад. $\Delta Q$ как раз равно разности собственных значений $h(R)$ при $R\to \infty$ и при $R\to 0$.

    Обозначим заряд $\alpha$-частицы $z$ = 2, материнского ядра --- $Z$. В пределе больших расстояний $\alpha$-частицу можно рассматривать как источник кулоновского поля $U_\alpha (r)=ze/r$,  которое в области дочернего атома можно рассматривать как почти однородное. В окрестности произвольной точки $R_0$ его можно разложить в ряд Тейлора:
\bea
U_\alpha (R) = U_\alpha (R_0) +
\nabla U_\alpha (R_0)(\m R - \m R_0) = \nonumber  \\ = U_\alpha (R_0)
+  \m F_\alpha (R_0) (\m R - \m R_0)\,,
\label{uas}
\eea
где ${\mathbf  F_\alpha} (r) = -({ze^2}/{r^2})\m n$ --- напряженность поля $\alpha$-частицы,  $\m n = \m r / r$  --- единичный вектор. Первое слагаемое в (\ref{uas}) приводит к сдвигу энергии электронов во внешнем поле $-ZeU_\alpha (R)$.  Второе можно рассматривать как источник штарковского сдвига энергии терма, который квадратичен по полю \cite{lan}.
Поэтому асимптотическое выражение для электронной энергии терма можно записать в виде \cite{dem}
\be
\epsilon (R) =
-\frac{(Z - z)ze^2}R - \frac {\beta} {R^4} \,. \label{4al}
\ee
Последнее слагаемое в (\ref{4al}) представляет собой  поляризационный потенциал.
Таким образом, мы нормируем электронную часть энергии терма на ноль при $R\to\infty$.

      Для пояснения физического смысла написанных соотношений полезно рассмотреть обратный процесс слияния $\alpha$-частицы с ядром. Если вводить $\alpha$-частицу от бесконечности в ядро, то электронный потенциал будет совершать часть работы по преодолению отталкивающего ядерного кулоновского потенциала, ``ускоряя'' частицу. При нормировке  на ноль при $R\to\infty$, $|\epsilon(0)|$ как раз равно работе, которую требуется совершить дополнительно по перемещению $\alpha$-частицы от нуля в бесконечность при распаде, и работе, которую совершает электронная оболочка, ускоряя  обратный процессе слияния $\alpha$-частицы с ядром. Эта работа и есть то изменение тепловыделения в реакции, которое отличает альфа-распад голого ядра от распада в атоме:
\be \Delta Q = -\epsilon(0)\,.
\label{Qbn} \ee
С учетом (\ref{Qbn}), действие (\ref{eq17}) переходит в
\bea
S_{at} = \int_{r_1'}^{r_2'}\sqrt{2m[U_C(R)+ \epsilon (0) + \Delta \epsilon (R) -(Q_\alpha -\Delta Q_\alpha)]} =
\label{eq17a} \\ = \sqrt{2m[U_C(R)+ \Delta \epsilon (R) -Q_\alpha]}\,. \label{Sat}
\eea
Таким образом, в результате (\ref{Qbn}), величина $\Delta Q$ в действии (\ref{eq17a}) строго сокращается и не фигурирует в выражении для вероятности $\alpha$-распада. А входит только гораздо более слабая вариация потенциальной энергии благодаря минимальной перестройке электронной оболочки до того момента, когда $\alpha$-частица пересечет потенциальный барьер. Это влияние относительно слабо и справедливо обычно опускается в  теории $\alpha$-распада \cite{fur}. Основная перестройка электронной оболочки происходит на гораздо больших расстояниях, принадлежащих уже другой шкале, --- порядка размера атома, и не оказывает влияния на уже состоявшийся к этому моменту распад ядра.

      В модели ЗО альфа-частице приходится преодолевать энергию притяжения электронов $U_e(R)$, которая аппроксимируется их средним  полем в атоме $\phi(R)$:
\be
U_e(R) = ze\phi(R)\,.   \label{fi}
\ee
      На это так же расходуется энергия альфа-частицы. В отличие от (\ref{4al}),  потенциал $\phi(R)$ (\ref{fi}) спадает к нулю  при $R\to\infty$ как $-Ze/R$. Таким образом,  он не учитывает потерю двух электронов при альфа-распаде. Поляризационный потенциал также не учитывается в модели ЗО. Однако более существенная численная разница состоит в том, что при $R\to 0$ потенциал $U_e (R)$, оставаясь отрицательным, монотонно убывает до величины    $ U_e(0)$. Обозначим
\be
\Delta Q^\prime=-U_e(0)\,.  \label{Qat}
\ee
По определению, данная модель предполагает электронную оболочку неизменной при прохождении альфа-частицы, таким образом упуская из виду потерю альфа-частицей энергии, идущей на перестройку электронной оболочки. Поэтому всегда
\be
\Delta Q^\prime < \Delta Q \,.
\ee
Для расчета вероятности альфа-распада в модели ЗО необходимо в действии (\ref{BN}) добавить $U_e(R) = U_e(0) + \Delta U_e(R)$ и произвести замену $Q_\alpha \to Q_\alpha - \Delta Q$. В результате придем к действию
\bea
S^\prime_{at} = \int_{r_1'}^{r_2'}\sqrt{2m[U_C(R)+ U_e(0) + \Delta U_e(R) - (Q_\alpha - \Delta Q)]} =  \nonumber   \\
= \int_{r_1'}^{r_2'}\sqrt{2m[U_C(R) + \Delta U_e(R) - Q_\alpha +\Delta Q+ U_e(0))]} \,. \label{Spat}
\eea
В (\ref{Spat})  $\Delta U_e(R)$ в первом приближении можно сопоставить с  $\Delta\epsilon (R)$ в (\ref{Sat}), что и выполнено в разделе \ref{comp}.  Помимо этого, однако, в модели ЗО в действие  (\ref{Spat})  дополнительно входит нефизическая величина $\Delta Q + U_e(0)$.

    Среди других физических ``нестыковок'' отметим, что ``замороженные'' на своих местах электроны должны бы были приводить к искривлению траектории альфа-частицы, что в свою очередь привело бы к подавлению альфа-распада, удлиняя путь под барьером. В действительности, из-за неравенства масс альфа-частицы и электронов траектория альфа-частицы прямолинейна, но она как бы увлекает за собой шлейф электронов, на что естественно должна затрачиваться энергия. В этом смысле можно сказать, что альфа-частица становится подобна мигдаловской квазичастице в ядре.

    С другой стороны, указанная выше некорректность асимптотического потенциала ЗО $\phi(R)$ в (\ref{fi}) на больших расстояниях  несущественна, так как относится к забарьерной области и, следовательно, не сказывается на проницаемость барьера. Однако она повод для размышлений, если кто-то захочет численно проверить учебники в отношении теории прохождения частицей барьера в поисках забарьерного отражения.   А
для интегрирования уравнения Шредингера во всем пространстве   необходимо прежде построить адиабатический потенциал, например методом самосогласованного среднего поля для двухцентровой задачи, переходящий на больших расстояниях в поляризационный. Справедливо замечено, что все нужное несложно, а все
сложное ---  ненужно \cite{skovoroda}.

\section{Результаты}
\subsection{Основной электронный терм }
\label{comp}

      Итак, в предыдущем разделе мы видели, что электронным термом называется уровень квазимолекулы,
рассматриваемый как функция расстояния между ядрами. Помимо отталкивающего кулоновского потенциала между ядрами, терм включает также энергию мгновенного электронного гамильтониана при фиксированных ядрах. Именно эта последняя энергия и должна быть добавлена к потенциальной кулоновской энергии ядер при вычислении действия для нейтрального атома.

В работе \cite{kar} эта часть терма, $\Delta \epsilon (R)$,   вычислена в первом порядке теории возмущений.  В этом приближении
\be
\Delta \epsilon^{(1)} (R)  = -ze^2 \sum_i \int \psi^2(r) \left( \frac 1{|\m r-\m R|} - \frac 1r \right)\ d^3r\,.    \label{eq4}
\ee
Суммирование в (\ref{eq4}) производится по всем занятым электронным состояниям атома, с учетом чисел заполнения. Вычисленный график $\Delta \epsilon (R)$  $^{226}$Ra приведен на рис. 1. Расчеты  проведены с релятивистскими волновыми функциями электронов. Для их вычисления использовался пакет программ RAINE \cite{RAINE}. Волновые функции  вычислены самосогласованным методом Дирака-Фока.

      Из рисунка следует, что на малых расстояниях $\epsilon (R)$  возрастает квадратично, тогда как традиционная теория межзвездной плазмы рассматривает электронные энергии постоянными в этой области (например, \cite{slptr}).  Учет этой зависимости приводит к наблюдаемым эффектам \cite{dzub}.

      Существенные для вероятности альфа-распада расстояния в области кулоновского барьера обычно составляют десятки ферми, что на два порядка меньше размера $K$-оболочки. Поэтому основной вклад в
$\Delta \epsilon (R)$  дают самые нижние оболочки с маленьким орбитальным моментом $l$, чьи волновые функции максимальны в этой области. Прежде всего, это  $K$-, $L_1$- и, с учетом релятивистских эффектов, $L_2$-оболочка.
Расчет показывает, что доминирующий вклад происходит от электронов на $K$-оболочке, ближайших к ядру.
При характерном расстоянии $R$ = 40 фм он составил
 81 процент. Относительный вклад от   $L_1$- и  $L_2$-оболочек соответственно оказался
13 и 1.4 процента. Для сравнения, на том же рисунке представлен график  $\Delta \epsilon_{\text{n.r.}} (R)$, вычисленный с нерелятивистскими волновыми и функциями Шредингера, с учетом только двух $K$-электронов. Разница огромная, особенно на малых расстояниях.  При $R\approx$  10 фм она составляет порядок величины. Даже в забарьерной области при $R\approx$  200 фм она сохраняется на уровне четырех раз.

    Причиной столь большого расхождения являются релятивистские эффекты, весьма существенные для тяжелых атомов. Сравнение дираковских и шредингеровских волновых функций представлено на рис. 2. Здесь представлены рассчитанные большие компоненты радиальных дираковских волновых функций $g(r)$ атома $^{226}$Ra и кулоновской волновой функции Шредингера с тем же $Z$ = 88. Как и ожидалось, разница наиболее существенна вблизи ядра. Релятивистские кулоновские функции расходятся в начале координат \cite{lan4}. Конечные  размеры ядра делают конечными в нуле и релятивистские волновые функции. Тем не менее, разница остается очень значительной.

\subsection{Самосогласованность адиабатического метода}

Другим важным  преимуществом адиабатического подхода является его самосогласованность. Она заключается в том, что  изменение потенциальной энергии альфа-частицы вдоль траектории от $R$ = 0 к $R\to \infty$ в точности равно работе, совершаемой пондермоторными силами, исходящими от электронной оболочки, при перемещении  альфа-частицы от бесконечно-удаленной точки в ядро, что в свою очередь в точности совпадает с изменением $\Delta Q$ (\ref{Qbn}). Ситуация иная в модели ЗО. Типичная разница между $|U_e(0)|$ и $\Delta Q$ составляет $\sim$50 эВ \cite{pol}. В свою очередь, как видно из рис. 1, эта величина составляет почти 100\% всего  изменения электронного терма  в области кулоновского барьера.

    На выражение в (\ref{eq4}) можно взглянуть иначе. Здесь первое слагаемое --- потенциальная энергия $\alpha$-частицы, находящейся в точке $\m R$, в поле, созданном электронной оболочкой. Из него вычитается второе слагаемое --- та же потенциальная энергия $\alpha$-частицы, находящейся в начале координат при $R = 0$. Следовательно,
\be
\Delta \epsilon^{(1)} (R)  = \Delta U_e(R) \,.  \label{eq4a}
\ee
Поэтому в первом порядке теории возмущений единственная разница с моделью ЗО состоит в наличии добавки $\Delta Q + U_e(0)$ в (\ref{Spat}).

Этот ложный вклад никак нельзя ликвидировать в модели ЗО математически корректным способом. Если его занулить принудительно, ``руками'', то, как показано в работе \cite{pol}, эффект экранирования уменьшается приблизительно втрое (треугольники на рис. 2 в работе \cite{pol}). При этом для $^{214}$Rn получается тот же результат, что и в работе \cite{kar} в адиабатическом подходе --- как и следовало ожидать, ввиду сказанного выше. Тем не менее, даже такое совпадение оправдано только в первом порядке теории возмущений, который  дает верхнюю границу эффекта. Более точный расчет квазимолекулярных состояний, за пределами первого порядка, мог бы следовательно привести и к значительно меньшей величине эффекта.

    В целом, можно утверждать, что поляризация электронной оболочки, производимая пролетающей альфа-частицей, смягчает эффект экранировки, что и приводит к на порядок  меньшей величине эффекта, вычисленного в адиабатическом подходе, по сравнению с моделью ЗО.
Попытка помочь модели с помощью формулы Хеллмана-Фейнмана \cite{pol} не помогает. Данная формула относится к нейтральным системам с $Z = N_e$, где $N_e$ --- число электронов. Последнее, однако, меняется в альфа-распаде. Более того, теорема Фейнмана справедлива только для устойчивых систем \cite{fey}, что очевидно не относится к ``замороженной'' оболочке. Можно заключить, что наличие слагаемого $\Delta Q + U_e(0)$ в (\ref{Spat}) остается главным дефектом модели ЗО, неустранимым в рамках  модели.

\subsection{Величина эффекта}

      Проводя расчеты по формуле (\ref{Pal}) с действием (\ref{BN}) для голого ядра и с действием (\ref{eq17}) в случае нейтрального атома, соответственно, найдем соответствующие вероятности туннелирования $ P$ и  $P_\text{at}$.  Их отношение определяет искомый эффект экранирования:
\be
 Y = (P/P_\text{at})-1  \,.
\label{Y} \ee
Значения $Y$ для различных альфа-распадчиков  приведены в таблице. Они в среднем порядка 10$^{-3}$ и в несколько раз меньше вычисленных в модели ЗО \cite{pol}, в соответствии со сказанным выше.
\begin{table}
\caption{Results for the relative change of the half-periods in
bare nuclides (last column)}
\begin{center}
\begin{tabular}{||c|c|c|c||}
 \hline \hline
Nuclide     &  Q, MeV  &  $T_{1/2}$&      $Y$, \%  \\
 \hline \hline
$^{144}_{60}$Nd     &   1.905  &  2.29$\times 10^{15}$ yr   &    0.24   \\
$^{214}_{86}$Rn&    9.208  &  0.27 $\mu$s   & 0.02\\
$^{226}_{88}$Ra&    4.871   & 1600 yr   &   0.23    \\
$^{252}_{98}$Cf&     6.217  &  2.645 yr &   0.28    \\
$^{241}_{99}$Es&   8.320    &    9 s        &   0.12    \\
$^{294}$118 &   11.65   &   0.89 ms &   0.27    \\
 \hline \hline
\end{tabular}
\end{center}
\label{altab}
\end{table}

\section{Заключение}

Современная экспериментальная техника позволяет произвести сравнение периодов $\alpha$-распада в обычных атомах и в голых ядрах, лишенных атомной оболочки. Возможное отличие представляет большой интерес, так как формирование и вылет $\alpha$-частицы из ядра происходят на расстояниях, очень малых по атомной шкале. Поэтому традиционно теория ядерных реакций строится без учета электронной оболочки. Данная работа продолжает теоретическое исследование механизма и ожидаемой величины эффекта. Для эксперимента существенно опираться на наиболее совершенную теорию. В этом отношении результаты, полученные в настоящей работе, представляют несомненный интерес.  Показано, что модель ЗО, применявшаяся на протяжении последних десятилетий, завышает величину эффекта в несколько раз вследствие того, что не учитывает процесса перестройки электронной оболочки при пролете сквозь нее $\alpha$-частицы и  сопряженную с ней потерю энергии $\alpha$-частицей.

      Адекватный теоретический подход дается адиабатическим методом, который как раз основан на учете указанных эффектов. Расчеты проведены в первом порядке теории возмущений, достаточном для описания при расстояниях $\lesssim$100 фм в области кулоновского барьера. Исследованы важные для планирования эксперимента подробности.

1) Показано, что несмотря на предсказуемо малую величину эффекта $\sim 10^{-3}$, имеется благоприятное для эксперимента обстоятельство: $\sim$80\% эффекта создается  $K$-электронами. Поэтому для наблюдения эффекта достаточно сравнить периоды распада в голом ядре с периодами в  одно- и (или) двухэлектронных атомах,   а вовсе не надо привлекать нейтральные атомы --- их накопление в кольцах ускорителя невозможно,  а столь малые  эффекты  надежнее наблюдать по разностным измерениям на одной и той же установке. Таким образом, можно произвести сравнительные измерения в одних и тех же накопительных кольцах с голым ядром и ионами различной кратности, например гелие-подобными.

2) Хотя рассчитанная величина эффекта $\sim$в 3 раза меньше, чем ожидавшаяся на основе расчетов в модели ЗО, все же расчеты в первом приближении теории возмущений дают  \emph{верхнюю границу} эффекта. Поэтому повышение точности для расчета двухцентровых квазимолекулярных термов представляет актуальную задачу.

3) В рамках адиабатического подхода понятна причина относительного успеха модели ЗО и пределы ее применимости. Изменение потенциальной энергии в интеграле действия в этой модели такое же, как и в первом порядке теории возмущений в адиабатическом подходе. Поэтому область относительной применимости модели ЗО такая же, как и первого порядка теории возмущений, то есть при малых сравнительно с атомными расстояниях от альфа-частицы до ядра. В эту область входит и кулоновский барьер для типичных
$\alpha$-распадчиков. За пределами кулоновского барьера  потенциальная энергия в модели ЗО теряет физическое правдоподобие. Это обстоятельство сыграло бы роль при уменьшении энергии $\alpha$-распада.

4) Результаты настоящей работы интересны и для экспериментов $(n, \alpha)$, подобных \cite{gled}, с мэвными нейтронами. В таких экспериментах могла бы проявиться резонансная структура в спектре $\alpha$-частиц, предсказанная в работах \cite{sochi,jpal,jmp}. Формулу (\ref{Sat}) можно трактовать как прохождение $\alpha$-частицей того же самого барьера, что и для голого ядра, но с переменной энергией  $Q_\alpha  - \Delta \epsilon (R)$. В то же время из рис. 2 следует, что на протяжении барьера эта величина меняется приблизительно на 200 эВ. Данное изменение как раз сравнимо с шириной линий, предсказанных в работе \cite{jmp}, и может следовательно повлиять на их форму или интенсивность. Это влияние, принимая во внимание теорему Купманса \cite{kar}, может быть  обнаружено в сравнительном эксперименте на накопительных кольцах.

5) Адиабатический подход представляется адекватным методом и для описания обратных реакций синтеза. Поэтому он представляет интерес в исследованиях плазмы в лабораториях или в звездной среде. В связи с тем, что для подобных реакций характерны  гораздо более низкие энергии, то и эффект ожидается гораздо более сильным. В то же время адекватный расчет может потребовать выхода за рамки первого порядка теории возмущений. Полученные выше результаты будут интересны и полезны в соответствующих лабораториях: ВНИИЭФ Саров, NIF,  GSI и других.

\clearpage
\begin{center}
ПОДПИСИ К РИСУНКАМ
\end{center}

\begin{enumerate}
\item
Рисунок 1. Вклады в основной терм различных электронных конфигураций: полный расчет по Дираку---Фоку --- сплошная линия, и нерелятивистский расчет с учетом только $1s$-электронов --- штрих-пунктирная линия.

\item
Рисунок 2.  Радиальные волновые функции Дирака для $1s$-электрона, большая компонента, для $^{226}$Ra,  --- сплошная линия, в сравнении с нерелятивистской кулоновской волновой функцией --- штрих-пунктир.

\end{enumerate}

\clearpage

\begin{figure}[!b]
\centerline{ \epsfxsize=15cm\epsfbox{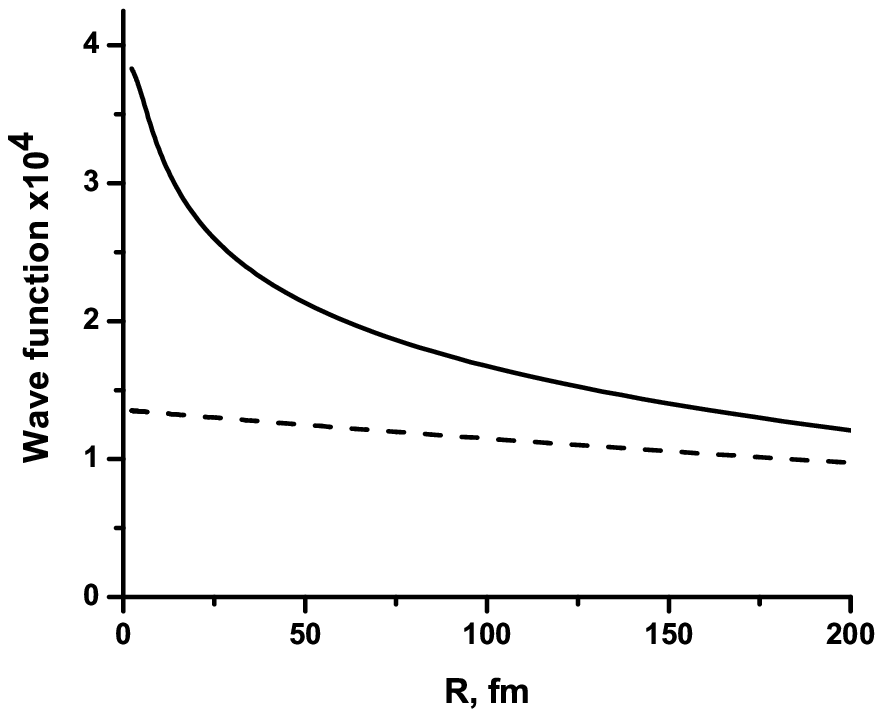}}
\caption{} \label{f1}
\end{figure}

\clearpage

\begin{figure}[!b]
\centerline{ \epsfxsize=15cm\epsfbox{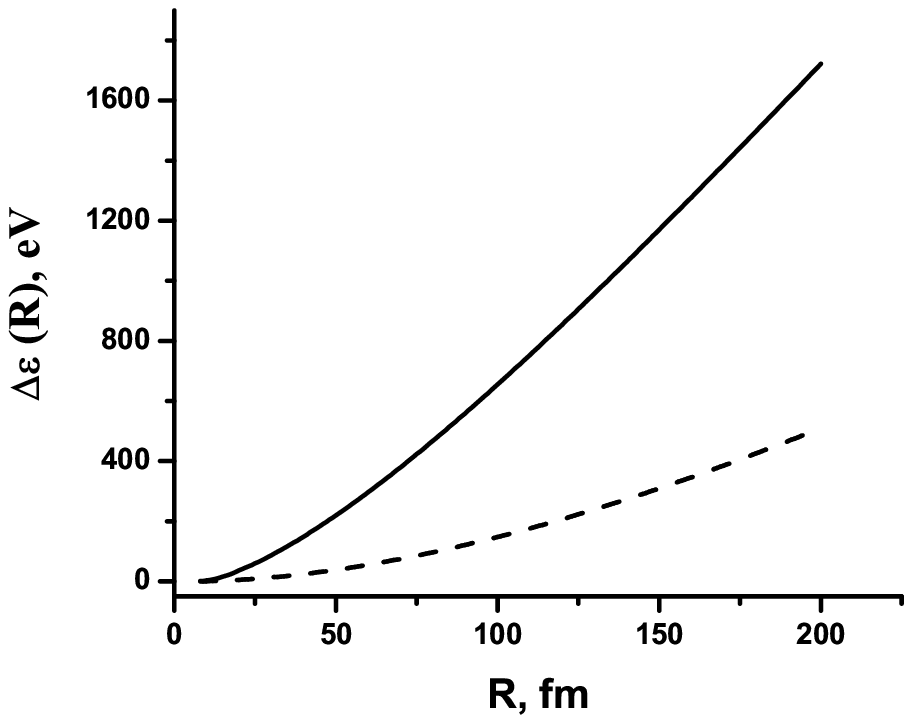}}
\caption{} \label{f2}
\end{figure}


\begin{thebibliography}{99}\fussy

\bibitem{kar}  F. F. Karpeshin, Phys. Rev. C {\bf 87}, 054319 (2013).

\bibitem{PR1} V. A. Erma, Phys Rev. {\bf 103}, 1784 (1952).

\bibitem{perl} W. Rubinson and M.L. Perlman, Phys. Lett. {\bf 40B}, 352 (1972).

\bibitem{PR2} K. U. Kettner et al., J. Phys. G: Nucl. Part. Phys. {\bf 32}, 489
(2006).

\bibitem{PRZin} N. T. Zinner, Nucl. Phys. A {\bf 781}, 81 (2007).


\bibitem{pol}  Z. Patyk, H. Geissel, Yu. A. Litvinov, A. Musumarra, and C. Nociforo, Phys. Rev. C {\bf 78}, 054317 (2008).

\bibitem{fey}  R. P. Feynman, Phys. Rev. C {\bf 56}, 340 (1939).

\bibitem{chu}  Yu.M. Tchuvilsky et al., IASEN 2013, South Africa, 2 -- 6 December 2013, PROGRAMME; ЯФ,  {\bf 76},  1537 (2013); ЯДРО -- 2012, 25 -- 30 июня 2012 г., Воронеж, Тезисы докладов, с. 63.
\bibitem{dzub} A.Ya. Dzublik, Phys. Rev. C {\bf 90}, 054619 (2014).
\bibitem{ober}  V. E. Oberacker, A. S. Umar, and F. F. Karpeshin, Prompt muon-induced fission: a sensitive probe for nuclear energy dissipation and fission dynamics,  in: Progress in Muon Research, ed. Frank Columbus, Nova Science Publishers, Inc., Hauppauge, NY, 2005,
http://arxiv.org/PS\_cache/nucl-th/pdf/0403/0403087.pdf

\bibitem{lea} {\it Leander P., Nilsson S.G., M\"{o}ller P.}, Phys. Lett.  {\bf B90}, 193 (1980).

\bibitem{risse} F. Risse, W. Bertl, P. David et al., Z. Phys. {\bf A339}, 427 (1991).

\bibitem{dem} Ю.Н.Демков, Д.Ф.Зарецкий, Ф.Ф.Карпешин, М.А.Листенгартен, В.Н.Островский, Письма в ЖЭТФ, {\bf 28}, 287, 1978; ЯФ,  {\bf 31}, 47, 1980.

\bibitem{berl} E.Ye.Berlovich, F.F.Karpeshin,  Phys. Lett. B, {\bf 177}, 260, 1986.

\bibitem{den} V. Yu. Denisov and H. Ikezoe, Phys. Rev. C {\bf 72}, 064613 (2005).

\bibitem{sochi} F. F. Karpeshin, in Exotic Nuclei EXON---2009, Proc. International Symposium,  27 September -- 3 October 2009, Sochi, Russia. AIP Conference Proceedings,   {\bf 1224}, 133,    Melville, New York, 2010.

\bibitem{fur} C. П.Кадменский, В.Фурман. Альфа-распад и родственные ядерные реакции. М.: Энергоатомиздат, 1985.

\bibitem{lan} {\it Ландау Л.Д., Лифшиц Е.М.}  Квантовая механика. Нерелятивистская теория, М.: Наука, 1974.

\bibitem{skovoroda} См. например,  Г. С. Сковорода, http://www.zitata.com/skovoroda.shtml










\bibitem {RAINE} I. M. Band, M. B. Trzhaskovskaya, C. W. Nestor Jr.,
       P. O. Tikkanen, S. Raman,   Atom. Data and Nucl. Data Tables {\bf 81}, 1 (2002); I. M. Band and  M. B. Trzhaskovskaya,  {\it ibid.} {\bf 55}, 43 (1993); {\bf 35}, 1 (1986).

\bibitem {slptr} E. E. Salpeter, Australian J. Phys. {\bf 7}, 373  (1954).

\bibitem{lan4} В.Б.Берестецкий, Е.М.Лившиц, Л.П.Питаевский, {\it Квантовая электродинамика}, М.: Физматлит, 2002.
\bibitem {gled} Yu.M. Gledenov,  G. Zhang, G. Khuukhenkhuu {\em et al.}, Phys. Rev. C {\bf 82}, 014601 (2010).
\bibitem {jpal} F. F. Karpeshin, G. LaRana, E. Vardaci, A. Brondi, R. Moro,
    S. N. Abramovich and V. I. Serov, J. Phys. G: Nucl. Part. Phys.
    {\bf 34}, 587 (2007).
\bibitem {jmp} F. F. Karpeshin, Intern. J. Modern Phys. {\bf B22}, 4709 (2008).

\end{thebibliography}
\end{document}